\documentclass[12pt]{article}

\usepackage{amssymb}
\usepackage{amsmath}
\usepackage{amscd}
\usepackage{latexsym}

\usepackage{graphicx}
\usepackage{array}

\usepackage{cite}

\newcommand{\be}{\begin{equation}}
\newcommand{\ee}{\end{equation}}
\newcommand{\vp}{\varphi}

\newcommand{\bt}{\beta}
\newcommand{\ra}{\rightarrow}

\newcommand{\Dlt}{\Delta}
\newcommand{\br}{{\bf r}}
\newcommand{\ba}{{\bf a}}
\newcommand{\bb}{{\bf b}}
\newcommand{\dlt}{\delta}
\newcommand{\prt}{\partial}
\newcommand{\om}{\omega}

\newcommand{\rgl}{\rangle}
\newcommand{\lgl}{\langle}

\begin{document}

{\parindent=0pt
Moscow Univ. Phys. Bull. {\bf 31}, 10--15 (1976). }

\vskip 2cm

\begin{center}

{\Large {\bf Theory of Perturbations with a Strong Interaction} \\ [5mm]
V.I.~Yukalov }  \\ [3mm]

{\it Department of Theoretical Physics \\ 
Moscow State University, Moscow 119991, Russia}

\end{center}

\vskip 2cm

\begin{abstract}
The theory of perturbations is suggested for statistical systems in the absence 
of small interaction parameters. A new form is advanced for self-consistent
conditions defining the optimal parameters for trial Green functions in iterating 
nonlinear propagator equations. Superharmonic, semiharmonic, and pseudoharmonic 
approximations for a molecular crystal are considered as examples.
\end{abstract}

\vskip 2cm
\section*{Theory of Perturbations}

Let us consider a statistical system in the five-dimensional space $\{ y\}$.
$$
y = \{ x,t\} \; ;   \qquad x = \{ r,f\}\; ;   \qquad \br = \{ r_i \}\; ;
\qquad i = 1,2,3 \; ,
$$
where $f$ is a variable characterizing internal degrees of freedom or macroscopic
state indices $[1,2]$ (or their combination). For brevity, we shall write
\be
\label{1}
\vp(y_1\ldots y_n) \equiv \vp(1\ldots n) \; ; \qquad
dy_1 \ldots dy_n \equiv d(1\ldots n) \; .
\ee

We assume that the exact solution of the equation of motion
\be
\label{2}
\int G^{-1}(13) \; G(32)\; d(3) = \dlt(12)
\ee
is unknown, but the solution of some model problem
\be
\label{3}
\int G_0^{-1}(13) \; G_0(32)\; d(3) = \dlt(12)
\ee
can be found. Here and subsequently $G$ denotes a causal Green's function, which 
is defined in the usual manner through field operators, or axiomatically [3]. 
Introducing the kernel
\be
\label{4}
K(12) = \int G_0(13) \; [ G_0^{-1}(32) - G^{-1}(32) ] \; d(3) \; ,
\ee
we can write, on the basis of (2) and (3), the Dyson equation
\be
\label{5}
G(12) = G_0(12) + \int K(13) G(32)  \; d(3) \; .
\ee

The nonlinear integral equation (5) can be solved by some approximate methods [4]. 
We shall, for example, integrate it following the scheme
$$
G_n \ra K_{n+1} \ra G_{n+1}  \; ,
$$
in which $K_n$ is the $n$ - iterated functional (4),
$$
K_{n+1} \equiv K\{ G^{-1}_{n+1} \} \; , \qquad G^{-1}_{n+1} \equiv  G^{-1}\{ G_n \} \; .
$$

Obviously, if the particle interaction is sufficiently strong, then the propagator of 
free particles cannot be taken as a zero approximation. One should take as $G_0$ a trial 
Green function, whose parameters are to be defined from additional conditions. The following 
self-consistent optimal conditions are suggested: the average values of some operators 
${\cal O}$, corresponding to observable quantities, calculated in the $n+1$ -th and $n$ -th 
iterations coincide, i.e.,
\be
\label{6}
\int \lim_{x_2\ra x_1}\; \lim_{t_{12}\ra -0} \;
{\cal O}(1) [ G_{n+1}(12) - G_n(12) ] \; dx_1 = 0 \; .
\ee
The expression $t_{12}$ implies the difference $t_1-t_2$. It is necessary to take as many
different equalities (6) as there is the number of trial parameters contained in $G_0$. Of 
all possible operators, one should select those ${\cal O}$, the averages of which have to be 
known with the greatest accuracy in the given problem. When the left-hand side of (6) is 
proportional to the average number or particles, then in taking thermodynamic limit one should 
divide Eq. (6) by
$$
N = \pm i \int \lim_{x_2\ra x_1}\; \lim_{t_{12}\ra -0} \; G(12) \; dx_1 \; .
$$

Stopping at the first iteration step and making use of (5), we get, instead of (6),
\be
\label{7}
\int \lim_{x_2\ra x_1}\; \lim_{t_{12}\ra -0} \;
{\cal O}(1) K_1(13) G_0(32) \; d(3) dx_1 = 0 \; .
\ee

Condition (7) can be employed for both equilibrium and nonequilibrium systems. In the 
case of the former it is better to use Fourier representation with respect to $\tau\equiv t-t'$,
$$
\vp(y,y') = \frac{1}{2\pi} \int \vp(x,x',\om) e^{-i\om\tau}\; d\om \; ,
$$
where the definite integral over the interval $(-\infty,+\infty)$ is assumed. Here
$$
K(x,x',\om) = \int G_0(x,x'',\om) [ G_0^{-1}(x'',x',\om) - G^{-1}(x'',x',\om) ]\; dx'' \; ,
$$
$$
N = \pm \; \frac{i}{2\pi} \int \lim_{x'\ra x}\; \lim_{\tau\ra +0} G(x,x',\om) 
e^{i\om\tau} \; d\om dx \; ,
$$
while Eq. (7) becomes
\be
\label{8}
\lim_{\tau\ra +0} \int G_0(x',x,\om) {\cal O}(x) K_1(x,x',\om) e^{i\om\tau}\; d\om dx dx' 
=  0 \; .
\ee
The operator ${\cal O}(x)$ is the same as ${\cal O}(y)$.

\section*{Self-Consistent Conditions} 

Let us see how the self-consistent conditions (8) are simplified in certain particular cases. 
Let $G$ be expanded in wave functions
\be
\label{9}
G_0(x,x'\om) =  \sum_n G_n(\om) \psi_n(x) \psi^*_n(x') \; .
\ee

The subscript $n$ in this expression means the total set of quantum indices characterizing 
the wave functions. In what follows, the upper sign is to be taken for Bose systems and the 
lower sign is to be used for Fermi systems:
$$
G_n(\om) = \frac{1\pm n(\om_n)}{\om-\om_n+i0} \mp \frac{n(\om_n)}{\om-\om_n-i0}\; ,
$$
$$
n(\omega) = \left(\exp(\beta \omega) \mp 1 \right)^{-1} \; .
$$

Using the notations
$$
\Dlt(x,x',\om) = G_0^{-1}(x,x',\om) - G_1^{-1}(x,x',\om) \; ,
$$
$$
\Dlt_{mn}(\om) = \int \psi^*_m(x) \Dlt(x,x',\om) \psi_n(x') \; dx dx' \; ,
$$
\be
\label{10}
{\cal O}_{mn} = \int \psi^*_m(x) {\cal O}(x) \psi_n(x) \; dx  \; ,
\ee
we transform Eq. (8) to
\be
\label{11}
\lim_{\tau\ra+0} \int e^{i\om\tau} \sum_{mn} G_m(\om) G_n(\om) 
{\cal O}_{mn} \Dlt_{nm}(\om) \; d\om = 0 \; .
\ee
Now $N=\sum_n n(\om_n)$. In converting from (8) to (11) it must be remembered that
$$
\lim_{x'\ra x} G(x,x',\om) = \pm \lim_{x'\ra x} G(x',x,-\om) \; .
$$

We assume that $\Dlt(x,x',\om)$ is not a function of frequency $\om$,
\be
\label{12}
\Dlt(x,x',\om) = \Dlt(x,x') \; ,
\ee
consequently, also $\Dlt_{mn}(\om) \equiv \Dlt_{mn}$ is not a function of $\om$. Noting 
that
$$
\lim_{\tau\ra+0} \int e^{i\om\tau} G_m(\om) G_n(\om) \; d\om =
\pm 2\pi i \; \frac{n(\om_m)-n(\om_n)}{\om_n-\om_m} \; ,
$$
we find from (11)
\be
\label{13}
\sum_{mn} {\cal O}_{mn} \; \frac{n(\om_m)-n(\om_n)}{\om_n-\om_m} \; \Dlt_{nm} = 0 \; .
\ee

If we assume that eigenfunctions of operator ${\cal O}(x)$ coincide with $\psi_n(x)$
$$
{\cal O}(x) \psi_n(x) = {\cal O}_n \psi_n(x) \; ,
$$
then ${\cal O}_{mn}=\dlt_{mn}{\cal O}_n$, and since
$$
\lim_{m\ra n} \; \frac{n(\om_m)-n(\om_n)}{\om_n-\om_m} = \bt n^2(\om_n) e^{\bt\om_n} \; ,
$$
equation (13) reduces to
\be
\label{14}
\sum_n {\cal O}_n n^2(\om_n) e^{\bt\om_n} \Dlt_{nn} = 0 \; .
\ee

It should be noted that the limiting transitions $m\ra n$, $\tau\ra +0$ for the integral
$$
I_{mn}(\tau) = \int e^{i\om\tau} G_m(\om) G_n(\om) \; d\om
$$
are noncommutative:
$$
\lim_{m\ra n}\; \lim_{\tau\ra+0} \; I_{mn}(\tau) = 
\pm 2\pi i n^2(\om_n)  e^{\bt\om_n} \; , \qquad
\lim_{\tau\ra+0} \; \lim_{m\ra n}\;  I_{mn}(\tau) = \mp \infty \; .
$$

Only the first sequence of limits is meaningful, which is natural, since initially
one specifies the equation of motion, also fixing $\Dlt(x,x',\om)$, and only then  
the operators ${\cal O}(x)$ are selected for self-consistent conditions.

Here it is particularly important to use causal Green functions. When (12) applies,
Eq. (11) for retarded Green functions becomes an identity due to the fact that for then
$\lim_{\tau\ra+0} I_{mn}(\tau)\equiv 0$.

If $n(\om_n)$ decreases rapidly with the increase in $n$, then it is possible to 
retain a finite number of terms in summation (14). Retaining the first term with 
$n=0$, we obtain the self-consistent condition
\be
\label{15}
\int \psi^*_0(x) \Dlt(x,x') \psi_0(x') \; dx dx' = 0 \; .
\ee

\section*{Example of Application}

Let us consider the application of the suggested techniques for an example of a nonmagnetic 
localized crystal, for which the Green's function is represented as a sum over the 
lattice sites $G=\sum G_a$), and when a pseudo-Hartree (Hartree approximation with an 
effective potential) equations [5,6] can be used. A classical crystal in the Hartree 
approximation has been treated in [7]). 

In the considered case, Eq. (10) becomes
$$
\Dlt(\br,\br',\om) = [ V_1(\br) - V_0(\br) ] \dlt(\br-\br') \; .
$$

The average field
$$
V_1(\br) = \int \rho(\br') \sum_\ba \Phi(\br,\br'+\ba) \; d\br' \; ,
$$
where $\ba$ is a lattice vector and $\Phi$ is a pseudopotential taking into account pair 
particle correlations. As is known [8], taking for $\Phi$ an interaction potential of bare 
particles most frequently produces divergences. 

For a zero approximation, we can accept the potential of the isotropic harmonic oscillator
$$
V_0(\br) = u_0 + \frac{m_0\om_0^2}{2} \; ( \br - \ba )^2 \; ; \qquad
u_0 = w \sum_\bb \Phi(\ba,\bb) \; .
$$
Here $w$ is the number of particles per elementary cell:
$$
w = \int \rho(\br) \; d\br = \sum_{nlm} n(\om_{nl}) \; , \qquad
\rho(\br) = \sum_{nlm} n(\om_{nl}) |\psi_{nlm}(\br) |^2 \; ,
$$
$$
\om_{nl} = u_0 + \om_0\left ( 2n +l + \frac{3}{2} \right ) - \mu\; .
$$

An anharmonic oscillator [9,10] could be taken for a starting approximation, however, 
this would have rapidly resulted in unjustified complication of calculations and hence 
it is unreasonable.

To describe the single trial parameter $\om_0$ we shall use the self-consistency condition (14)
with the unit operator ${\cal O}$:
\be
\label{16}
\sum_{nlm} n^2(\om_{nl}) e^{\bt\om_{nl}} \Dlt_{nlm} = 0 \; ,
\ee
\be
\label{17}
\Dlt_{nlm} = \int |\psi_{nlm}(\br - \ba)|^2 [ V_1(\br) - V_0(\br) ] \; d\br \; .
\ee
The average
$$
\lgl \; \vp(\br)\; \rgl_{nm} = \int |\psi_{nlm}(\br )|^2 \vp(\br) \; d\br
$$
for $\vp=\br^2$ can be easily found from the virial theorem, which yields
$$
\lgl \; \br^2 \; \rgl_{nlm} = \left ( 2n + l + \frac{3}{2} \right ) 
\frac{1}{m_0\om_0} \; .
$$
Since $n(\om_{nl})$ falls off rapidly with increasing $n$ and $l$, it is possible 
to retain a single term from (16) with $n=l=0$, which is equivalent to (15). Then We 
find for $\om_0$
$$
\om_0 = \frac{4}{3} \; (\Phi_0 - u_0) \; , \qquad
\Phi_s = w \sum_\bb | \psi_0(\br) |^2 \; | \psi_0(\br') |^2 \; 
\Phi(\br+\ba,\br'+ \bb) \; d\br d\br' \; ,
$$
where
$$
w = n ( E_0 - \mu ) \; , \qquad E_0 = u_0 + \frac{3\om_0}{2} \; .
$$

As far as the trial frequency $\om_0$ is defined through the total potential $\Phi$,
without involving its expansion in Taylor series, that is, with allowance for the 
anharmonicity of all orders, without exception, this approximation can be termed  
{\it superharmonic}. Usually, however, the interaction potential is expanded in series 
in terms of the deviations from the lattice sites.

The principal contribution to integral (17) is made by the region near $\br=\ba$, 
and it is hence sensible to use the expansion
$$
\Phi(\br+\ba,\br'+\bb) = \sum_{n=0}^\infty \frac{1}{n!} \; 
\prt^n_{ab}(\br,0) \Phi(\ba,\br'+\bb) \; , 
$$
\be
\label{18}
\prt_{ab}(\br,\br') = \sum_{i=1}^3 \left ( r_i\; \frac{\prt}{\prt a_i} + r_i'\;
\frac{\prt}{\prt b_i} \right ) \; .
\ee

Using the second order of the above expansion and employing to the properties
$$
\int r_i \rho(\br) \; d\br = 0 \; , \qquad 
\int r_i r_j \rho(\br) \; d\br = \dlt_{ij} \int r_i^2 \rho(\br) \; d\br \; , 
$$
$$
\sum_{m=-l}^{+l} \lgl \; r_i^2\; \rgl_{nlm} = 
\frac{1}{3} \; \lgl \; \br^2 \; \rgl_{nlm} \; ,
$$
and the notation
$$
\Phi_h = w \sum_\ba \lgl \; \Phi(\ba,\br+\bb) \; \rgl_0 \; , \qquad
\om_h^2 = \frac{w}{3m_0} \sum_\ba \sum_{i=1}^3 \frac{\prt^2}{\prt a_i^2} \; 
\lgl \; \Phi(\ba,\br+\bb) \; \rgl_0 \; ,
$$
we find from (15) the trial frequency
$$
\om_0 = \frac{1}{3} \; \left | \; 2(\Phi_h - u_0) + 
\sqrt{4(\Phi_h-u_0)^2 + 9\om_h^2} \; \right |
$$
as the frequency in the semiharmonic approximation. Retaining the $n+1$ terms in
expansion (18), we would have obtained an $n-$-th order semiharmonic approximation.

If we carry out expansion (17) with the consideration of the fact that the function
$\psi_{nlm}(\br-\ba)$ decreases fast with the distance from the lattice site $\br=\ba$,
and expand the pseudopotential, then expanding it in both coordinates, we get
\be
\label{19}
\Phi(\br+\ba,\br'+\bb) = 
\sum_{n=0}^\infty \frac{1}{n!} \; \prt^n_{ab} \Phi(\ba,\bb) \; .
\ee
We introduce the notation
$$
\om_1^2 = \frac{w}{3m_0} \sum_\bb \sum_{i=1}^3 \frac{\prt^2}{\prt a_i^2}\; \Phi(\ba,\bb) \; ,
\qquad
\om_2^2 = \frac{w}{3m_0} \sum_\bb \sum_{i=1}^3 \frac{\prt^2}{\prt b_i^2}\; \Phi(\ba,\bb) \; ,
$$
$$
r_0^2 = \frac{1}{w} \sum_{nl} n(\om_{nl}) \lgl \; \br^2\; \rgl_{nlm} \; .
$$

Equation (16) in the second order of expansion (19) contains the quantity
$$
\frac{2}{m_0} \sum_{nl} \Dlt_{nlm} = \left [ \; \om_1^2 + (2l+1) \om_0^2 \; \right ]
\lgl \; \br^2\; \rgl_{nlm} + ( 2l + 1) \om_2^2 r_0^2 \; .
$$

This approximation, which is identical to the pseudoharmonic in the simplest case of 
$n=l=0$, when
$$
r_0^2 = \lgl \; \br^2\; \rgl_0 = \frac{3}{2m_0} \; \om_0 \; ,
$$ 
yields
\be
\label{20}
\om_0 =\sqrt{\om_1^2 + \om_2^2 } \; .
\ee

Incorporating  $n+1$ terms in (19) would have resulted in an $n-$- th order pseudoharmonic 
approximation.

For the potential $\Phi(\br,\br')=\Phi(\br-\br')$, since $\om_1=\om_2$, then on the 
basis of (20), we would have $\om_0 = \om_1\sqrt{2}$.

\vskip 1cm

{\parindent=0pt
{\it Submitted May 15, 1974 \\
Department of Theoretical Physics} }


\vskip 2cm


\begin{thebibliography}{99}
\bibitem{1}
V.I. Yukalov, Moscow Univ. Phys. Bull. {\bf 25}, 49 (1970). 

\bibitem{2}
V.I. Yukalov, Moscow Univ. Phys. Bull. {\bf 26}, 22 (1971).

\bibitem{3}
V.I. Yukalov, Theor. Math. Phys. {\bf 17}, 1244 (1973).

\bibitem{4}
{\it Approximate Solutions of Operator Equations} (Nauka, Moscow, 1969).

\bibitem{5}
V.I. Yukalov, Moscow Univ. Phys. Bull. {\bf 27}, 59 (1972).

\bibitem{6}
V.I. Yukalov, Proc. PFU Phys. {\bf 59}, 91 (1972).

\bibitem{7}
V.I. Zubov and Y.P. Terletsky, Ann. Physik {\bf 24}, 97 (1970).

\bibitem{8}
S.V. Tyablikov, J. Exp. Theor. Phys. {\bf 20}, 16 (1950).

\bibitem{9}
C.M. Bender and T.T. Wu, Phys. Rev. D {\bf 7}, 1620 (1973).

\bibitem{10}
P. Lu, S. Wald, and B. Young, Phys. Rev. D  {\bf 7}, 1701 (1973).

\end{thebibliography}
\end{document}